\documentclass[11pt]{article}
\usepackage{amssymb,amsmath,amsfonts}
\usepackage{graphicx}
\usepackage{graphics}
\usepackage{eepic,epsfig}

\begin{document}

\title{Polarization of the conformal vacuum in the Milne universe}
\author{T. A. Petrosyan \\
\\
\textit{Department of Physics, Yerevan State University,}\\
\textit{1 Alex Manoogian Street, 0025 Yerevan, Armenia}}
\maketitle

\begin{abstract}
The vacuum expectation values (VEVs) of the field squared and
energy-momentum tensor for a massless scalar field are investigated in the
Milne universe with general number of spatial dimensions. The vacuum state
depends on the choice of the mode functions in the canonical quantization
procedure and we assume that the field is prepared in the conformal vacuum.
As the first step an integral representation for the difference of the
Wightman functions corresponding to the conformal and Minkowski vacua is
derived. The mean field squared and energy-momentum tensor are obtained in
the coincidence limit. It is shown that the Minkowski vacuum state is
interpreted as a thermal one with respect to the conformal vacuum. The
thermal factor is of the Bose-Einstein type in odd dimensional space and of
the Fermi-Dirac type in even number of spatial dimensions.
\end{abstract}


\bigskip

\section{Introduction}

It is well-known that the vacuum state in quantum field theory, in general,
depends on the choice of the mode functions used in the canonical
quantization procedure. If the Bogoliubov $\beta $-coefficient relating two
sets of modes is different from zero then the corresponding vacuum states
are inequivalent. An example, widely considered in the literature (see \cite%
{Birr82,Full96,Cris08} and references therein), is provided by the Minkowski
and Fulling-Rindler vacua in flat spacetime. The Minkowski vacuum
corresponds to the modes of inertial observer and it preserves all the
symmetries of the Minkowski spacetime (is maximally symmetric). The
Fulling-Rindler vacuum is realized in the quantization procedure based on
the modes of uniformly accelerated observers. For those observers horizons
are present corresponding to the light cones that divide the spacetime into
four sections. The right and left patches are referred to as the right and
left Rindler wedges, respectively. The upper and lower patches are covered
by the Milne coordinates and are referred to as the Milne universe.\ 

The line element of the flat spacetime in the Milne coordinates is of the
Friedmann-Robertson-Walker type with a scale factor being a linear function
of the corresponding time coordinate. The spacetime is foliated by negative
curvature spatial sections. The corresponding geometry serves as a simple
model for the investigation of quantum-field-theoretical effects in
time-dependent backgrounds. A relatively large number of problems are
exactly solvable and motivated by that various aspects of the dynamics of
quantum fields in the Milne universe have been discussed in \cite{Somm74}-%
\cite{Saha20}. In the present paper we consider the local characteristics of
the conformal vacuum state (referred to as the C-vacuum) for a massless
scalar field in the Milne universe.

The paper is organized as follows. In the next section we present the normal
modes and the Wightman functions for the C- and Minkowski vacua in the case
of a massive scalar field with general curvature coupling parameter. The
general results are specified for a massless field in section \ref%
{sec:WFconf}. The expression for the mean field squared and energy-momentum
tensor are discussed in section \ref{sec:VEV}. The main results are
summarized in section \ref{sec:Conc}.

\section{Wightman functions in conformal and Minkowski vacua}

\label{sec:Wig}

The line element for the $(D+1)$-dimensional Milne universe is expressed as%
\begin{equation}
ds^{2}=dt^{2}-t^{2}(dr^{2}+\sinh ^{2}rd\Omega _{D-1}^{2}),  \label{ds2}
\end{equation}%
where for the time and dimensionless radial coordinates one has $0\leq
t<\infty $, $0\leq r<\infty $, and $d\Omega _{D-1}^{2}$ is the line element
on a $(D-1)$-dimensional sphere. The spatial part corresponds to a constant
negative curvature space covered by the hyperspherical coordinates $%
(r,\vartheta ,\phi )$. For the set of angular coordinates we have $\vartheta
=(\theta _{1},\theta _{2},\ldots \theta _{n})$, $0\leq \theta
_{k}\leq \pi $, $k=1,2,\ldots ,n$ and $0\leq \phi \leq 2\pi $%
, where $n=D-2$. Note that the spacetime described by the line element (\ref%
{ds2}) is flat. That is explicitly seen introducing new coordinates $(T,%
\mathbf{R})$, with $\mathbf{R}=(R,\vartheta ,\phi )$, in accordance with 
\begin{equation}
T=t\cosh r,\;R=t\sinh r.  \label{TR}
\end{equation}%
The line element (\ref{ds2}) takes the Minkowskian form $%
ds^{2}=dT^{2}-dR^{2}-R^{2}d\Omega _{D-1}^{2}$ in hyperspherical spatial
coordinates. As seen from (\ref{TR}), $T>R$ and the Milne coordinates $%
(t,r,\vartheta ,\phi )$ cover the patch of the Minkowski spacetime inside
the future light cone. In the region corresponding to the past light cone we
take $T=t\cosh r$, $R=-t\sinh r$ with the time coordinate $-\infty <t\leq 0$%
. The remaining regions of the Minkowski spacetime, $R>|T|$, correspond to
the Rindler patches.

Passing to new coordinates $(\eta ,\bar{r})$, $-\infty <\eta <\infty $, in
accordance with 
\begin{equation}
t=ae^{\eta /a}\equiv ae^{\bar{\eta}},\;r=\bar{r}/a,  \label{Coord}
\end{equation}%
where $a$ is a constant with dimension of length, the line element (\ref{ds2}%
) is presented in a conformally-static form%
\begin{equation}
ds^{2}=\left( t/a\right) ^{2}\left[ d\eta ^{2}-d\bar{r}^{2}-a^{2}\sinh
^{2}\left( \bar{r}/a\right) d\Omega _{D-1}^{2}\right] .  \label{ds2n}
\end{equation}%
The line element in the square brackets describes a static spacetime with a
constant negative curvature space. The curvature radius of the latter is
determined by $a$ and in the limit $a\rightarrow \infty $ for fixed $\eta $
and $\bar{r}$, from (\ref{ds2n}) the Minkowskian line element in spherical
coordinates is obtained. For two spacetime points $(T^{\prime },\mathbf{R}%
^{\prime })$ and $(T,\mathbf{R})$ the spacetime interval between them is
expressed as%
\begin{equation}
\left( \Delta T\right) ^{2}-\left\vert \Delta \mathbf{R}\right\vert
^{2}=2tt^{\prime }\left( \cosh \Delta \bar{\eta}-\cosh \zeta \right) ,
\label{cr1}
\end{equation}%
where $\Delta T=T^{\prime }-T$, $\Delta \mathbf{R}=\mathbf{R}^{\prime }-%
\mathbf{R}$, $\Delta \bar{\eta}=\bar{\eta}^{\prime }-\bar{\eta}$, and $\zeta 
$ is defined by the relation%
\begin{equation}
\bar{u}=\cosh r\cosh r^{\prime }-\sinh r\sinh r^{\prime }\cos \theta =\cosh
\zeta .  \label{zeta}
\end{equation}%
Here, $\theta $ is the angle between directions $\left( \vartheta ^{\prime
},\phi ^{\prime }\right) $ and $\left( \vartheta ,\phi \right) $. In the
special case $\theta =0$, corresponding to points on the same radial
directions, we have $\zeta =r^{\prime }-r\equiv \Delta r$.

Having described the background geometry we turn to the field content. We
consider a quantum scalar field $\varphi (x)$ with curvature coupling
parameter $\xi $. The field equation reads%
\begin{equation}
\left( \Box +m^{2}+\xi R\right) \varphi =0,  \label{Feq}
\end{equation}%
where $\Box $ is the d'Alembert operator and for background under
consideration the Ricci scalar is zero, $R=0$. Let $\{\varphi _{\sigma
}\left( x\right) ,\varphi _{\sigma }^{\ast }\left( x\right) \}$ be the
complete set of mode functions obeying the field equation and specified by
the quantum numbers $\sigma $. For the modes corresponding to the C-vacuum $%
\sigma =\left( z,m_{p}\right) $ and 
\begin{equation}
\varphi _{\sigma }\left( x\right) =N_{\sigma }\frac{J_{-iz}(mt)}{t^{(D-1)/2}}%
\frac{P_{iz-1/2}^{1-D/2-l}(\cosh r)}{\sinh ^{D/2-1}r}Y(m_{p};\vartheta ,\phi
),  \label{phi}
\end{equation}%
where $J_{\nu }\left( x\right) $ is the Bessel function, $P_{\rho }^{\gamma
}(x)$ is the associated Legendre function of the first kind and $%
Y(m_{p};\vartheta ,\phi )$ are the hyperspherical harmonics. In (\ref{phi}), 
$0\leq z<\infty $ and $m_{p}=(l,m_{1},\ldots ,m_{n})$ with $l=0,1,2,\ldots $%
. For the integers $m_{1},m_{2},\ldots ,m_{n}$ one has $-m_{n-1}\leqslant
m_{n}\leqslant m_{n-1}$ and $0\leqslant m_{n-1}\leqslant m_{n-2}\leqslant
\cdots \leqslant m_{1}\leqslant l$. The coefficient $N_{\sigma }$ is
determined by the standard normalization condition for the equation (\ref%
{Feq}) and is determined from%
\begin{equation}
\left\vert N_{\sigma }\right\vert ^{2}=\frac{z\left\vert \Gamma \left(
(D-1)/2+l+iz\right) \right\vert ^{2}}{2N\left( m_{p}\right) }.  \label{C}
\end{equation}%
The factor $N\left( m_{p}\right) $ comes from the normalization condition
for the hyperspherical harmonics and its explicit form will not be required
in the discussion below.

The correlations of the vacuum fluctuations of quantum fields at different
spacetime points $x$ and $x^{\prime }$ are determined by the two-point
functions. Here we consider the Wightman function. The latter can be
evaluated by using the mode-sum formula 
\begin{equation}
W(x,x^{\prime })=\sum_{\sigma }\varphi _{\sigma }\left( x\right) \varphi
_{\sigma }^{\ast }\left( x^{\prime }\right) .  \label{WF}
\end{equation}%
The expression of the Wightman function for the conformal vacuum (denoted
here as $W_{\text{\textrm{C}}}(x,x^{\prime })$) can be obtained by using the
corresponding formula for the Wightman function from \cite{Saha20}:%
\begin{eqnarray}
W_{\text{\textrm{C}}}(x,x^{\prime }) &=&\frac{\left( tt^{\prime }\right)
^{(1-D)/2}}{2nS_{D}}\sum_{l=0}^{\infty }\frac{\left( 2l+n\right)
C_{l}^{n/2}(\cos \theta )}{(\sinh r\sinh r^{\prime })^{D/2-1}}%
\int_{0}^{\infty }dy\,y|\Gamma (l+\left( D-1\right) /2+iy)|^{2}  \notag \\
&&\times J_{-iy}(mt)J_{iy}(mt^{\prime })P_{iy-1/2}^{-\mu
}(u)P_{iy-1/2}^{-\mu }(u^{\prime }),  \label{wfcv}
\end{eqnarray}%
where $S_{D}=2\pi ^{D/2}/\Gamma \left( D/2\right) $, $C_{l}^{n/2}(\cos
\theta )$ is the Gegenbauer polynomial, and 
\begin{equation}
u=\cosh r,\,u^{\prime }=\cosh r^{\prime }.
\end{equation}

By using the addition theorem for the associated Legendre functions from 
\cite{Henr55} (for the correction of the missprint see \cite{Saha21}) it can
be shown that 
\begin{eqnarray}
&&\overset{\infty }{\underset{l=0}{\sum }}\left( l+\frac{n}{2}\right)
C_{l}^{n/2}\left( \cos \theta \right) \left\vert \Gamma \left( l+\left(
D-1\right) /2+iy\right) \right\vert ^{2}P_{iy-1/2}^{-l-n/2}\left( u\right)
P_{iy-1/2}^{-l-n/2}\left( u^{\prime }\right)  \notag \\
&&\qquad =\frac{2^{-n/2}}{\Gamma \left( n/2\right) }\left[ \left(
u^{2}-1\right) \left( u^{\prime 2}-1\right) \right] ^{n/4}\left\vert \Gamma
\left( \left( D-1\right) /2+iy\right) \right\vert ^{2}\frac{%
P_{iy-1/2}^{-n/2}\left( \bar{u}\right) }{\left( \bar{u}^{2}-1\right) ^{n/4}},
\label{AdP2}
\end{eqnarray}%
with $\bar{u}$ given by (\ref{zeta}). By using this result, the expression (%
\ref{wfcv}) is further simplified as 
\begin{equation}
W_{\text{\textrm{C}}}(x,x^{\prime })=\frac{\left( tt^{\prime }\right)
^{(1-D)/2}}{2\left( 2\pi \right) ^{D/2}}\int_{0}^{\infty
}dy\,yJ_{-iy}(mt)J_{iy}(mt^{\prime })\left\vert \Gamma \left( iy+\left(
D-1\right) /2\right) \right\vert ^{2}\frac{P_{iy-1/2}^{-n/2}\left( \bar{u}%
\right) }{\left( \bar{u}^{2}-1\right) ^{n/4}}.  \label{wfcv1}
\end{equation}%
The VEVs of the field squared and energy-momentum tensor are obtained from
two-point functions in the coincidence limit. That limit is divergent and a
renormalization is required. In the present paper we are interested in the
difference of the local characteristics of the C- and Minkowski vacua. The
latter is obtained from the difference $\Delta W(x,x^{\prime })=W_{\text{%
\textrm{C}}}(x,x^{\prime })-W_{\text{\textrm{M}}}(x,x^{\prime })$, where%
\begin{equation}
W_{\text{\textrm{M}}}(x,x^{\prime })=\frac{m^{\left( D-1\right) /2}}{\left(
2\pi \right) ^{\left( D+1\right) /2}}\frac{K_{\left( D-1\right) /2}(m\sqrt{%
\left\vert \Delta \mathbf{R}\right\vert ^{2}-\left( \Delta T\right) ^{2}})}{%
\left[ \left\vert \Delta \mathbf{R}\right\vert ^{2}-\left( \Delta T\right)
^{2}\right] ^{\left( D-1\right) /4}},  \label{wfm3}
\end{equation}%
is the Wightman function for the Minkowski vacuum. Here, we assume that $%
\left\vert \Delta \mathbf{R}\right\vert >|\Delta T|$. Note that, in
accordance with (\ref{cr1}), this corresponds to $|\zeta |>|\Delta \bar{\eta}%
|$. The expressions in the other regions of the Minkowski spacetime are
obtained by the analytical continuation. The local geometry for both the
C-vacuum in the Milne patch and for the Minkowski vacuum is the same and,
hence, the difference $\Delta W_{\text{\textrm{C}}}(x,x^{\prime })$ is
finite and can be directly used for the evaluation of the local VEVs.

\section{Wightman function for a massless field}

\label{sec:WFconf}

The expression (\ref{wfcv1}) for the Wightman function is further simplified
for a massless field. In the limit $m\rightarrow 0$ for the product of the
Bessel functions one has 
\begin{equation}
J_{-iy}(mt)J_{iy}(mt^{\prime })\approx \frac{\sinh \left( \pi y\right) }{\pi
y}e^{iy\Delta \bar{\eta}},
\end{equation}%
and the formula takes the form%
\begin{equation}
W_{\text{\textrm{C}}}(x,x^{\prime })=\frac{\left( tt^{\prime }\right)
^{(1-D)/2}}{\left( 2\pi \right) ^{D/2+1}}\int_{0}^{\infty }dy\,\sinh \left(
\pi y\right) e^{iy\Delta \bar{\eta}}\left\vert \Gamma \left( iy+\frac{D-1}{2}%
\right) \right\vert ^{2}\frac{P_{iy-1/2}^{-n/2}\left( \bar{u}\right) }{%
\left( \bar{u}^{2}-1\right) ^{n/4}}.  \label{Wconf}
\end{equation}%
By using the properties of the associated Legendre function this expression
can be rewritten as%
\begin{eqnarray}
W_{\text{\textrm{C}}}(x,x^{\prime }) &=&\left( -1\right) ^{q}\frac{\left(
tt^{\prime }\right) ^{(1-D)/2}}{\left( 2\pi \right) ^{D/2+1}}\partial _{\bar{%
u}}^{q}\int_{0}^{\infty }dy\,\sinh \left( \pi y\right) e^{iy\Delta \bar{\eta}%
}  \notag \\
&&\times \left\vert \Gamma \left( \frac{D-1}{2}-q+iy\right) \right\vert ^{2}%
\frac{P_{iy-1/2}^{1-D/2+q}\left( \bar{u}\right) }{\left( \bar{u}%
^{2}-1\right) ^{(D-2q-2)/4}}.  \label{Wconf2}
\end{eqnarray}%
with $q$ being a non-negative integer. In the massless limit, by taking into
account the relation (\ref{cr1}), for the Minkowski vacuum one gets 
\begin{equation}
W_{\text{\textrm{M}}}(x,x^{\prime })=\frac{\left( tt^{\prime }\right)
^{\left( 1-D\right) /2}}{2\left( 2\pi \right) ^{\left( D+1\right) /2}}\frac{%
\Gamma \left( \left( D-1\right) /2\right) }{\left( \cosh \zeta -\cosh \Delta 
\bar{\eta}\right) ^{\left( D-1\right) /2}}.  \label{Wcm2}
\end{equation}%
For the further transformation we will consider the odd and even values for
the spatial dimension $D$ separately.

For even values of spatial dimension, taking $q=D/2-1$, the Wightman
function is expressed as%
\begin{equation}
W_{\text{\textrm{C}}}(x,x^{\prime })=-\frac{\left( tt^{\prime }\right)
^{(1-D)/2}}{2\left( -2\pi \right) ^{D/2}}\left( \frac{\partial _{\zeta }}{%
\sinh \zeta }\right) ^{D/2-1}\int_{0}^{\infty }dy\,\tanh \left( \pi y\right)
e^{iy\Delta \bar{\eta}}P_{iy-1/2}\left( \cosh \zeta \right) .  \label{wfce}
\end{equation}%
The corresponding function for the Minkowski vacuum is presented in the form 
\begin{equation}
W_{\text{\textrm{M}}}(x,x^{\prime })=-\frac{\left( tt^{\prime }\right)
^{\left( 1-D\right) /2}}{2^{3/2}\left( -2\pi \right) ^{D/2}}\left( \frac{%
\partial _{\zeta }}{\sinh \zeta }\right) ^{D/2-1}\frac{1}{\sqrt{\cosh \zeta
-\cosh \Delta \bar{\eta}}}.  \label{wfmed1}
\end{equation}%
For the evaluation of the difference of the Wightman functions it is
convenient to present $W_{\text{\textrm{M}}}(x,x^{\prime })$ in an integral
form 
\begin{equation}
W_{\text{\textrm{M}}}(x,x^{\prime })=-\frac{\left( tt^{\prime }\right)
^{\left( 1-D\right) /2}}{2\left( -2\pi \right) ^{D/2}}\left( \frac{\partial
_{\zeta }}{\sinh \zeta }\right) ^{D/2-1}\int_{0}^{\infty }dy\,\cos \left(
y\Delta \bar{\eta}\right) P_{iy-1/2}\left( \cosh \zeta \right) .
\label{wfmed2}
\end{equation}%
Here we have used the relation \cite{Prud2}%
\begin{equation}
\int_{0}^{\infty }dy\,\cos \left( yv\right) P_{iy-1/2}\left( w\right) =\frac{%
1}{\sqrt{2}\sqrt{w-\cosh v}},  \label{IntForm}
\end{equation}%
valid for $w>\cosh v$. Thus, for the difference of the Wightman functions we
get%
\begin{equation}
\Delta W(x,x^{\prime })=\frac{\left( tt^{\prime }\right) ^{(1-D)/2}}{2\left(
-2\pi \right) ^{D/2}}\int_{0}^{\infty }dy\left[ \frac{2\cos \left( y\Delta 
\bar{\eta}\right) }{e^{2\pi y}+1}-i\tanh \left( \pi y\right) \sin \left(
y\Delta \bar{\eta}\right) \right] \frac{P_{iy-1/2}^{D/2-1}\left( \bar{u}%
\right) }{(\bar{u}^{2}-1)^{(D-2)/4}},  \label{Wceven}
\end{equation}%
where the relation $\partial _{\bar{u}}^{D/2-1}P_{iz-1/2}\left( \bar{u}%
\right) =(\bar{u}^{2}-1)^{(2-D)/4}P_{iz-1/2}^{D/2-1}\left( \bar{u}\right) $
was used.

Now we pass to the odd values of $D$. For this case in (\ref{Wconf2}) we
take $q=(D-3)/2$ and the associated Legendre function in the integrand of (%
\ref{Wconf2}) is reduced to the function $P_{iy-1/2}^{-1/2}\left( \bar{u}%
\right) $. The latter is expressed in terms of elementary functions and for $%
\left( \Delta \bar{\eta}\right) ^{2}\neq $ $\zeta ^{2}$ one gets 
\begin{equation}
W_{\text{\textrm{C}}}(x,x^{\prime })=\frac{\left( tt^{\prime }\right)
^{(1-D)/2}}{2\left( -2\pi \right) ^{\left( D+1\right) /2}}\left( \frac{%
\partial _{\zeta }}{\sinh \zeta }\right) ^{\left( D-3\right) /2}\frac{2\zeta
/\sinh \zeta }{\zeta ^{2}-\left( \Delta \bar{\eta}\right) ^{2}},
\label{wfco1}
\end{equation}%
where $\zeta $ is defined in accordance with (\ref{zeta}). For a massless
field and for odd values of $D$ the Wightman function (\ref{Wcm2}) for the
Minkowski vacuum is presented as

\begin{equation}
W_{\text{\textrm{M}}}(x,x^{\prime })=\frac{\left( tt^{\prime }\right)
^{\left( 1-D\right) /2}}{2\left( -2\pi \right) ^{\left( D+1\right) /2}}%
\left( \frac{\partial _{\zeta }}{\sinh \zeta }\right) ^{\left( D-3\right) /2}%
\frac{1}{\cosh \zeta -\cosh \Delta \bar{\eta}}.  \label{wfmo}
\end{equation}%
Here we have used the relation (\ref{cr1}).\ For the difference of the
Wightman functions, that determines the difference in the local VEVs, we
obtain%
\begin{equation}
\Delta W(x,x^{\prime })=\frac{\left( tt^{\prime }\right) ^{(1-D)/2}}{2\left(
-2\pi \right) ^{\left( D+1\right) /2}}\left( \frac{\partial _{\zeta }}{\sinh
\zeta }\right) ^{\left( D-3\right) /2}\left[ \frac{2\zeta /\sinh \zeta }{%
\zeta ^{2}-\left( \Delta \bar{\eta}\right) ^{2}}-\frac{1}{\cosh \zeta -\cosh
\Delta \bar{\eta}}\right] .  \label{Wdifo}
\end{equation}%
This representation is well adapted for the evaluation of the differences
between local VEVs of the field squared and energy-momentum tensor.

In the discussion above we have considered the difference in the Wightman
functions for the C- and Minkowski vacua. Similar expressions are obtained
for the differences of other two-point functions. In particular the VEVs of
the field squared are obtained from the Hadamard function $G(x,x^{\prime })$%
. For odd $D$ the corresponding expression is obtained from (\ref{Wdifo})
with an additional coefficient 2. In the case of even $D$ from (\ref{Wceven}%
) for the difference $\Delta G(x,x^{\prime })=G_{\text{\textrm{C}}%
}(x,x^{\prime })-G_{\text{\textrm{M}}}(x,x^{\prime })$ we get%
\begin{equation}
\Delta G(x,x^{\prime })=\frac{2\left( tt^{\prime }\right) ^{(1-D)/2}}{\left(
-2\pi \right) ^{D/2}}\int_{0}^{\infty }dy\frac{\cos \left( y\Delta \bar{\eta}%
\right) }{e^{2\pi y}+1}\frac{P_{iy-1/2}^{D/2-1}\left( \bar{u}\right) }{(\bar{%
u}^{2}-1)^{(D-2)/4}}.  \label{Gceven}
\end{equation}

\section{VEV of the field squared}

\label{sec:VEV}

As a local characteristic of the C-vacuum first let us consider the VEV of
the field squared. It is obtained from the Hadamard function in the
coincidence limit as $\Delta \left\langle \varphi ^{2}\right\rangle
=\lim_{x^{\prime }\rightarrow x}\Delta G(x,x^{\prime })/2$. If we
renormalize the corresponding VEV for the Minkowski vacuum to zero, $%
\left\langle \varphi ^{2}\right\rangle _{\mathrm{M}}^{\mathrm{(ren)}}=0$,
then $\Delta \left\langle \varphi ^{2}\right\rangle $ gives the renormalized
VEV for the C-vacuum, $\Delta \left\langle \varphi ^{2}\right\rangle
=\left\langle \varphi ^{2}\right\rangle _{\mathrm{C}}^{\mathrm{(ren)}}$. For
even values of $D$ we use the relation (see, for example, \cite{Nist10})%
\begin{equation}
\lim_{\bar{u}\rightarrow 1}\frac{P_{iy-1/2}^{D/2-1}\left( \bar{u}\right) }{%
\left( \bar{u}^{2}-1\right) ^{(D-2)/4}}=\frac{2^{1-D/2}\Gamma \left(
iy+\left( D-1\right) /2\right) }{\Gamma (D/2)\Gamma \left( iy-(D-3)/2\right) 
}.  \label{RelPl}
\end{equation}%
The expression for the VEV can be transformed to the form%
\begin{equation}
\Delta \left\langle \varphi ^{2}\right\rangle =-\frac{2^{-D}t^{1-D}}{\pi
^{D/2+1}\Gamma (D/2)}\int_{0}^{\infty }dy\,e^{-\pi y}\left\vert \Gamma
\left( iy+\left( D-1\right) /2\right) \right\vert ^{2}.  \label{phi2ev}
\end{equation}%
Alternatively, by using the properties of the gamma function we can see that%
\begin{equation}
\Delta \left\langle \varphi ^{2}\right\rangle =-\frac{2(4\pi )^{-D/2}}{%
\Gamma (D/2)t^{D-1}}\int_{0}^{\infty }dy\,\frac{y^{D-2}A_{D}\left( y\right) 
}{e^{2\pi y}+1},  \label{phi2ev2}
\end{equation}%
where for even $D$ we have defined%
\begin{equation}
A_{D}\left( y\right) =\prod\limits_{l=0}^{D/2-2}\left[ \left( l+1/2\right)
^{2}/y^{2}+1\right] .  \label{ADev}
\end{equation}%
As seen, the VEV is always negative.

For odd values of $D$ it is convenient firstly to put $\Delta \bar{\eta}=0$, 
$\theta =0$ in the expression (\ref{Wdifo}). With this choice we have $\zeta
=\Delta r$. The VEV of the field squared is presented as%
\begin{equation}
\Delta \left\langle \varphi ^{2}\right\rangle =-\frac{b_{D}t^{1-D}}{12(2\pi
)^{\left( D+1\right) /2}}\,,  \label{phi2odd}
\end{equation}%
where the coefficient $b_{D}$ is defined by the relation%
\begin{equation}
b_{D}=6(-1)^{\frac{D-1}{2}}\lim_{\bar{u}\rightarrow 1}\partial _{\bar{u}%
}^{\left( D-3\right) /2}\left[ \frac{2\left( \bar{u}^{2}-1\right) ^{-1/2}}{%
\text{\textrm{arccosh}}\left( \bar{u}\right) }-\frac{1}{\bar{u}-1}\right] .
\label{bD}
\end{equation}%
In particular,%
\begin{equation}
b_{3}=1,\;b_{5}=\frac{11}{30},\;b_{7}=\frac{191}{630},\;b_{9}=\frac{2497}{%
6300}.  \label{bD1}
\end{equation}%
It is of interest to note that the expression (\ref{phi2odd}) can also be
written in the form%
\begin{equation}
\Delta \left\langle \varphi ^{2}\right\rangle =-\frac{2(4\pi )^{-D/2}}{%
\Gamma (D/2)t^{D-1}}\int_{0}^{\infty }dy\,\frac{y^{D-2}A_{D}\left( y\right) 
}{e^{2\pi y}-1},  \label{phi2odd2}
\end{equation}%
where for odd $D$%
\begin{equation}
A_{D}\left( y\right) =\prod\limits_{l=0}^{\left( D-3\right) /2}\left(
l^{2}/y^{2}+1\right) .  \label{ADodd}
\end{equation}%
As before, the VEV is negative. We can combine the expressions for even and
odd values of the spatial dimension in a single formula 
\begin{equation}
\Delta \left\langle \varphi ^{2}\right\rangle =-\frac{B_{D}}{t^{D-1}},
\label{phi22}
\end{equation}%
where%
\begin{equation}
B_{D}=\frac{2(4\pi )^{-D/2}}{\Gamma (D/2)}\int_{0}^{\infty }dy\,\frac{%
y^{D-2}A_{D}\left( y\right) }{e^{2\pi y}+\left( -1\right) ^{D}}.  \label{BD}
\end{equation}

\section{VEV of the energy-momentum tensor}

\label{sec:EMT}

In this section we will consider the VEV of the energy-momentum tensor.
Through the Einstein semiclassical equations it determines the backreaction
of quantum effects on the background geometry. By taking into account that
for the background under consideration the Ricci tensor is zero, the
difference in the VEVs for the C- and Minkowski vacua, $\Delta \left\langle
T_{ik}\right\rangle =\left\langle T_{ik}\right\rangle _{\text{\textrm{C}}%
}-\left\langle T_{ik}\right\rangle _{\mathrm{M}}$, is evaluated on the basis
of the formula%
\begin{equation}
\Delta \left\langle T_{ik}\right\rangle =\frac{1}{2}\underset{x^{\prime
}\rightarrow x}{\lim }\partial _{i^{\prime }}\partial _{k}\Delta G\left(
x,x^{\prime }\right) +\left[ \left( \xi -\frac{1}{4}\right) g_{ik}\Box -\xi
\nabla _{i}\nabla _{k}\right] \Delta \left\langle \varphi ^{2}\right\rangle ,
\label{TikVev}
\end{equation}%
where $\nabla _{i}$ is the covariant derivative operator and $\xi $ is the
curvature coupling parameter. If we renormalize the VEV for the Minkowski
vacuum to zero, then $\Delta \left\langle T_{ik}\right\rangle $ gives the
mean energy-momentum tensor for the C-vacuum. Relations between the separate
components follow from the symmetry of the problem and from general
relations. First of all, the spatial geometry is isotropic and the stresses
are equal, $\Delta \left\langle T_{1}^{1}\right\rangle =\Delta \left\langle
T_{2}^{2}\right\rangle =\cdots =\Delta \left\langle T_{D}^{D}\right\rangle $%
. Next, by taking into account that all the components depend on the time
coordinate only, from the covariant continuity equation $\nabla
_{k}\left\langle T_{i}^{k}\right\rangle =0$ we get%
\begin{equation}
\Delta \left\langle T_{1}^{1}\right\rangle =\frac{1}{Dt^{D-1}}\partial
_{t}\left( t^{D}\Delta \left\langle T_{0}^{0}\right\rangle \right) .
\label{conteq}
\end{equation}%
Additionally, one has the trace relation%
\begin{equation}
\Delta \left\langle T_{i}^{i}\right\rangle =D(\xi -\xi _{D})\Box \Delta
\left\langle \varphi ^{2}\right\rangle ,  \label{Trace}
\end{equation}%
where $\xi _{D}=(D-1)/(4D)$ is the curvature coupling parameter for a
conformally coupled field. By taking into account that $\Delta \left\langle
\varphi ^{2}\right\rangle \propto 1/t^{D-1}$ it is easy to see that $\Box
\Delta \left\langle \varphi ^{2}\right\rangle =0$ and, hence, the VEV of the
energy-momentum tensor is traceless. This leads to the relation $\Delta
\left\langle T_{0}^{0}\right\rangle =-D\Delta \left\langle
T_{1}^{1}\right\rangle $ between the energy density and the stresses. From
this relation and from (\ref{conteq}) we get%
\begin{equation}
\Delta \left\langle T_{i}^{k}\right\rangle =\frac{C_{D}}{t^{D+1}}\text{%
\textrm{diag}}\left( 1,-1/D,\ldots ,-1/D\right) .  \label{Tstruc}
\end{equation}%
The problem is reduced to the evaluation of the constant $C_{D}$.

We will evaluate the component $\Delta \left\langle T_{11}\right\rangle $.
For the derivative in the last term of (\ref{TikVev}) one obtains%
\begin{equation}
\nabla _{1}\nabla _{1}\Delta \left\langle \varphi ^{2}\right\rangle =\left(
D-1\right) \Delta \left\langle \varphi ^{2}\right\rangle .  \label{Derphi}
\end{equation}%
For the evaluation of the first term we consider the cases of even and odd $%
D $ separately. For even $D$ the difference of the Hadamard functions is
given by (\ref{Gceven}). Taking $\theta =0$, $\Delta \eta =0$, in the
coincidence limit we get%
\begin{equation}
\frac{1}{2}\lim_{x^{\prime }\rightarrow x}\partial _{1^{\prime }}\partial
_{1}\Delta G\left( x,x^{\prime }\right) =-\frac{t^{1-D}}{\left( -2\pi
\right) ^{D/2}}\int_{0}^{\infty }dy\,\frac{1}{e^{2\pi y}+1}\lim_{\bar{u}%
\rightarrow 1}\partial _{\Delta r}^{2}\frac{P_{iy-1/2}^{D/2-1}\left( \cosh
\Delta r\right) }{\sinh ^{D/2-1}\Delta r}.  \label{lim1}
\end{equation}%
From the recurrence relations for the associated Legendre function the
following relation can be shown:%
\begin{equation}
\partial _{\Delta r}^{2}\frac{P_{iy-1/2}^{D/2-1}\left( \cosh \Delta r\right) 
}{\sinh ^{D/2-1}\Delta r}=\frac{P_{iy-1/2}^{D/2+1}\left( \cosh \Delta
r\right) }{\sinh ^{D/2-1}\Delta r}+\cosh \Delta r\frac{P_{iy-1/2}^{D/2}%
\left( \cosh \Delta r\right) }{\sinh ^{D/2}\Delta r}.  \label{RelP3}
\end{equation}%
The contribution of the first term in the right-hand side tends to zero in
the limit $\Delta r\rightarrow 0$ and we get%
\begin{equation}
\frac{1}{2}\lim_{x^{\prime }\rightarrow x}\partial _{1^{\prime }}\partial
_{1}\Delta G\left( x,x^{\prime }\right) =-\frac{2^{-D}\pi ^{-D/2-1}}{D\Gamma
(D/2)t^{D-1}}\int_{0}^{\infty }dy\,e^{-\pi y}\left\vert \Gamma \left(
iy+\left( D+1\right) /2\right) \right\vert ^{2}.  \label{lim2}
\end{equation}%
Substituting (\ref{Derphi}) and (\ref{lim2}) into (\ref{TikVev}) with $i=k=1$
and comparing with (\ref{Tstruc}) for even values of $D$ one finds%
\begin{eqnarray}
C_{D} &=&-\frac{\pi ^{-D/2-1}}{2^{D}\Gamma (D/2)}\int_{0}^{\infty
}dy\,e^{-\pi y}\left\vert \Gamma \left( iy+\left( D-1\right) /2\right)
\right\vert ^{2}\left[ y^{2}+D\left( D-1\right) \left( \xi _{D}-\xi \right) %
\right]  \notag \\
&=&-\frac{2^{1-D}\pi ^{-D/2}}{\Gamma (D/2)}\int_{0}^{\infty }dy\,\frac{%
y^{D-2}A_{D}\left( y\right) }{e^{2\pi y}+1}\left[ y^{2}+D\left( D-1\right)
\left( \xi _{D}-\xi \right) \right] ,  \label{CDeven}
\end{eqnarray}%
where $A_{D}(y)$ is given by (\ref{ADev}).

In the case of odd $D$, as in the case of the mean field squared, we can
take $\Delta \bar{\eta}=0$ and $\theta =0$. For the coincidence limit of the
derivative of the Hadamard function we get%
\begin{equation}
\lim_{x^{\prime }\rightarrow x}\partial _{1^{\prime }}\partial _{1}\Delta
G\left( x,x^{\prime }\right) =\frac{t^{1-D}}{\left( -2\pi \right) ^{\left(
D+1\right) /2}}\lim_{r^{\prime }\rightarrow r}\partial _{\Delta
r}^{2}\partial _{\bar{u}}^{\left( D-3\right) /2}\left[ \frac{1}{\bar{u}-1}-%
\frac{2\left( \bar{u}^{2}-1\right) ^{-1/2}}{\text{\textrm{arccosh}}\left( 
\bar{u}\right) }\right] .  \label{DGlim}
\end{equation}%
By taking into account that $\bar{u}=\cosh \Delta r$, we see that $\partial
_{\Delta r}^{2}=\bar{u}\partial _{\bar{u}}+(\bar{u}^{2}-1)\partial _{\bar{u}%
}^{2}$. The contribution of the last term vanishes in the limit $r^{\prime
}\rightarrow r$ ($\bar{u}\rightarrow 1$) and by using the definition (\ref%
{bD}) one obtains 
\begin{equation}
\lim_{x^{\prime }\rightarrow x}\partial _{1^{\prime }}\partial _{1}\Delta
G\left( x,x^{\prime }\right) =-\frac{b_{D+2}t^{1-D}}{6\left( 2\pi \right)
^{\left( D+1\right) /2}}.  \label{DGlim2}
\end{equation}%
This result with the combination of (\ref{phi2odd}) and (\ref{Derphi}) leads
to the following expression for the coefficient in (\ref{Tstruc}):%
\begin{equation}
C_{D}=D\frac{\xi \left( D-1\right) b_{D}-b_{D+2}}{12(2\pi )^{\left(
D+1\right) /2}}.  \label{CDn}
\end{equation}%
It can be checked that the latter expression, valid for odd values of $D$,
may also be written in the integral form%
\begin{equation}
C_{D}=-\frac{2^{1-D}}{\pi ^{D/2}\Gamma (D/2)}\int_{0}^{\infty }dy\,\frac{%
y^{D-2}A_{D}\left( y\right) }{e^{2\pi y}-1}\left[ y^{2}+D\left( D-1\right)
\left( \xi _{D}-\xi \right) \right] ,  \label{CDodd}
\end{equation}%
with $A_{D}(y)$ defined by (\ref{ADev}).

Let us compare the differences $\Delta \left\langle \varphi
^{2}\right\rangle $ and $\Delta \left\langle T_{ik}\right\rangle $ with the
differences in the corresponding VEVs between the Fulling-Rindler and
Minkowski vacua. The right Rindler wedge is covered by the coordinates $%
(\tau _{\mathrm{R}},\chi ,\mathbf{x}_{\mathrm{R}})$, with $-\infty <\tau _{%
\mathrm{R}}<+\infty $, $0\leq \chi <\infty $, $\mathbf{x}_{\mathrm{R}}=(x_{%
\mathrm{R}}^{2},\ldots ,x_{\mathrm{R}}^{D})$, and the corresponding line
element has the form 
\begin{equation}
ds_{\mathrm{M}}^{2}=\chi ^{2}d\tau _{\mathrm{R}}^{2}-d\chi ^{2}-d\mathbf{x}_{%
\mathrm{R}}^{2}.  \label{ds2Ri}
\end{equation}%
For a massless field, the difference of the mean field squared in the
Fulling-Rindler and Minkowski vacua is given by the expression \cite{Saha02}%
\begin{equation}
\left\langle \varphi ^{2}\right\rangle _{\mathrm{FR}}-\left\langle \varphi
^{2}\right\rangle _{\mathrm{M}}=-\frac{B_{D}}{\chi ^{D-1}},  \label{ph2FR}
\end{equation}%
where the coefficient $B_{D}$ is the same as in (\ref{BD}). The difference
in VEV of the energy-momentum tensor is expressed as (no summation over $i$)%
\begin{equation}
\left\langle T_{i}^{k}\right\rangle _{\mathrm{FR}}-\left\langle
T_{i}^{k}\right\rangle _{\mathrm{M}}=-\frac{2\delta _{i}^{k}(4\pi )^{-D/2}}{%
\Gamma (D/2)\chi ^{D+1}}\int_{0}^{\infty }dy\,\frac{y^{D-2}A_{D}\left(
y\right) }{e^{2\pi y}+(-1)^{D}}f_{0}^{(i)}(y),  \label{TikFR}
\end{equation}%
where%
\begin{eqnarray}
f_{0}^{(0)}(y) &=&-Df_{0}^{(1)}(y)=y^{2}+D(D-1)\left( \xi _{D}-\xi \right) ,
\notag \\
f_{0}^{(i)}(y) &=&-y^{2}/D+(D-1)^{2}\left( \xi _{D}-\xi \right)
,\;i=2,3,\ldots ,D.  \label{f0i}
\end{eqnarray}%
For a conformally coupled massless field one gets 
\begin{equation}
\left\langle T_{i}^{k}\right\rangle _{\mathrm{FR}}-\left\langle
T_{i}^{k}\right\rangle _{\mathrm{M}}=C_{D}\chi ^{-D-1}{\mathrm{diag}}\left(
1,-1/D,\ldots ,-1/D\right) ,
\end{equation}%
with the same coefficient as in (\ref{Tstruc}). For non-conformally coupled
fields the stresses for the Fulling-Rindler vacuum are anisotropic.

\section{Conclusion}

\label{sec:Conc}

The Milne universe is well suited for studying various aspects of the
influence of background geometry on the properties of quantum fields. In
particular, the study of properties of vacuum state is of special interest.
We have investigated the local properties of the C-vacuum for a massless
scalar field. Among the most important local characteristics are the
expectation values of the field squared and energy-momentum tensor. They are
obtained from the two-point functions in the coincidence limit of the
arguments. For the renormalization of the VEVs the subtraction of the
corresponding VEVs for the Minkowski vacuum is sufficient. This is related
to the fact that the Milne universe is flat and the divergences in the VEVs
for C- and Minkowski vacua are the same.

For a massless scalar field we have derived relatively simple
representations for the difference in the Wightman and Hadamard functions
for C- and Minkowski vacua. The renormalized mean field squared for C-vacuum
is directly obtained from the difference taking the coincidence limit. For
the evaluation of the renormalized mean energy-momentum tensor we have used
the formula (\ref{TikVev}). The mean field squared is given by (\ref{phi22})
with the coefficient (\ref{BD}). From the symmetry of the problem it follows
that VEV of the energy-momentum tensor should have the form (\ref{Tstruc}).
In order to obtain the expression for the coefficient $C_{D}$ we have
evaluated the 11-component. The coefficient is given by (\ref{CDeven}) for
even $D$ and by (\ref{CDodd}) in odd spatial dimensions. The formulas for
the mean field squared and energy-momentum tensor show that, from the point
of view of the quantization procedure in terms of the mode functions based
on the line element (\ref{ds2}), the Minkowski vacuum appears as a thermal
state. It is of interest to note that thermal factor is the Bose-Einstein
one for odd number of spatial dimensions and Fermi-Dirac type in even number
of spatial dimensions. We have emphasized that this feature is also present
in the relations between the VEVs in Fulling-Rindler and Minkowski vacua.
Similar features between the hyperbolic and Bunch-Davies vacua in de Sitter
spacetime have been discussed in \cite{Saha21a,Saha21}.

\section*{Acknowledgments}

I am indebted to Prof. Aram A. Saharian for constant support and valuable
advice. I am also grateful to Prof. Levon Sh. Grigoryan and Prof. Roland M.
Avagyan for encouraging discussions and productive colaboration.

\section*{Funding}

The work was supported by the Committee of Science of the Ministry of
Education, Science, Culture and Sport RA in the frames of the research
projects No. 20AA-1C005 and No. 21AG-1C047. The work was partly supported by
the "Faculty Research Funding Program 2020" (PMI Science and Enterprise
Incubator Foundation).

\end{document}